\documentclass[aps,prd,preprint,superscriptaddress,showpacs,preprintnumbers,amsmath,amssymb]{revtex4-1}
 \usepackage{booktabs}
\usepackage{graphicx}% Include figure files
\usepackage{epstopdf}
\usepackage{dcolumn}% Align table columns on decimal point
\usepackage{float}
\usepackage[normalem]{ulem}
\usepackage{color}
 \usepackage{multirow}
\usepackage{subfigure}
\usepackage{cancel}
\newcommand{\be}{\begin{equation}}

\newcommand{\ee}{\end{equation}}
\newcommand{\bea}{\begin{eqnarray}}

\newcommand{\eea}{\end{eqnarray}}

\newcommand{\pup}{p^\uparrow}

\newcommand{\pdown}{p^\downarrow}

\newcommand{\bfp}{\mbox{\boldmath $p$}}

\def\lsim{\mathrel{\rlap{\lower4pt\hbox{\hskip1pt$\sim$}}\raise1pt\hbox{$<$}}}
\def\gsim{\mathrel{\rlap{\lower4pt\hbox{\hskip1pt$\sim$}}\raise1pt\hbox{$>$}}}

\begin{document}

\title{Transverse Single Spin Asymmetry in 
$p+p^\uparrow \to  D +X $ \\}

\author{Rohini M. Godbole}
\email{rohini@chep.iisc.ernet.in}
\author{Abhiram Kaushik}
\email{abhiramb@chep.iisc.ernet.in}
\affiliation{Centre for High Energy Physics, Indian Institute of Science, Bangalore, India.}

\author{Anuradha Misra}
\email{misra@physics.mu.ac.in}

\affiliation{Department of Physics, University of Mumbai, Mumbai, India.}

\begin{abstract}

We present expected values of the single spin asymmetry (SSA) in $D$-meson production in  the process $p+p^\uparrow \rightarrow D^0 + X $  at RHIC energy $(\sqrt{s}=200\text{ GeV})$ using a generalized parton model (GPM) approach.  For this purpose, we use the  fits to the gluon Sivers function recently obtained by D'Alesio, Murgia and Pisano. We find that the peak asymmetry predictions for these lie in a broad range, $0.5\%\lesssim A_N\lesssim10\%$, for the kinematic regions considered. This is to be compared with the upper bound of 18\% expected for the maximal gluon Sivers function. We extend our analysis to two other centre of mass energies of proposed experiments - AFTER@LHC $(\sqrt{s}=115\text{ GeV})$ and a future RHIC run $(\sqrt{s}=500\text{ GeV})$. We further investigate (at $\sqrt{s}=200$ GeV) the effect of the transverse momentum dependent (TMD) evolution of the unpolarized parton distribution functions (PDFs) and of the gluon Sivers function on the  asymmetry predictions.  We find that inclusion of evolution causes an overall reduction of the asymmetry predictions.  For example, predictions for the peak asymmetry reduce by a factor of 3 or more for the case of the saturated gluon Sivers function. This decrease is similar to that noticed earlier for single spin asymmetry in the electroproduction of $J/\psi$.

%Valid PACS numbers may be entered using the \verb+\pacs{#1}+ command.
\end{abstract}

%\keywords{Suggested keywords}%
\maketitle

\section{\label{intro}Introduction}

Transverse momentum dependent (TMD)  parton distribution functions (PDFs) and fragmentation functions, collectively called TMDs, have been a subject of keen interest in the past decade on account of their role in understanding the large azimuthal and transverse single spin asymmetries (SSAs) observed in high energy processes. These asymmetries, which arise due to the orbital motion of quarks and gluons or due to recoil of gluons radiated off the active quarks, cannot be explained within the collinear factorization formalism  of  Quantum Chromodynamics (QCD). One needs to extend the  collinear QCD factorization scheme to include transverse motion  of partons. There are two theoretical approaches to generalize the conventional collinear  approach: (a) the  twist-3 collinear approach which is based on the inclusion of twist-3 quark gluon correlations within collinear factorization  and (b) the TMD  approach which is based on transverse momentum dependent PDFs and fragmentation functions. The twist-3 approach is suitable in a kinematic region where the transverse momentum of a particle observed in final state is large and is of the same order of magnitude as the other major scale of the process, whereas the TMD approach is valid  when the corresponding transverse momentum is  small.  However, in the intermediate region both approaches are applicable and  have been shown to be equivalent in the common region of validity \cite{Ji:2006ub, Koike:2007dg, Bacchetta:2008xw}

The TMD approach is based on a generalization of parton model, wherein one takes into account the transverse motion of partons by replacing the usual collinear partons distribution functions $f(x)$ by transverse momentum dependent PDFs $f(x, k_\perp)$, where $k_\perp$ is  the transverse momentum  and $x$  the longitudinal momentum fraction of the proton momentum that the parton carries. TMD factorization theorems have been proved rigorously only for SIDIS, DY and $e^+e^-$  annihilation processes. For processes for which TMD factorization has not been proved, a phenomenological approach, known as the generalized parton model (GPM) has been used to make predictions for SSAs. These include both two-scale processes such as $ep^\uparrow \rightarrow e+J/\psi +X$ \cite{Godbole:2013bca, Godbole:2014tha} and $p p^\uparrow \rightarrow jet +\pi +X$  \cite{D'Alesio:2010am}, and single-scale processes like $pp^\uparrow\rightarrow\pi+X$ \cite{D'Alesio:2004up} and $pp^\uparrow\rightarrow D+X$. We consider the latter here. It has been pointed out that the TMDs extracted using such a phenomenological approach may not be independent of the process used~\cite{pages:2015ika}. TMDs are sensitive to, among other things, initial state/final state interactions specific to the process under consideration \cite{Gamberg:2010tj}. Due to all these reasons, it is both interesting and important to try to extract the TMDs from a variety of processes and compare them. These measurements are sure to shed light on the multitude of issues involved.

One of the important TMDs is the Sivers function which arises due to the fact that the distribution of quarks and gluons in a transversely polarized hadron need not be symmetric about the axis of collision. This asymmetry in distribution, called the Sivers effect, causes an azimuthal asymmetry in the distribution of produced particles when one of the colliding particles is a transversely polarized hadron. The Sivers function, which parametrizes the correlation between the intrinsic motion of the partons and the transverse spin of the hadron,  can be considered as  the probability of finding an unpolarized parton inside a transversely polarized nucleon. There have been a number of phenomenological studies of the quark  Sivers functions, involving extraction of Sivers function parameters using polarised semi-inclusive deeply inelastic scattering (SIDIS) data from the HERMES, COMPASS and JLAB experiments~\cite{Anselmino-PRD72,kp09}. 

These initial fits~\cite{Anselmino-PRD72,kp09} were performed using the generalized parton model (GPM) framework where the complete QCD evolution of the TMDs was not taken into account. The transverse momentum dependence was factored out and the TMDs were modelled as a standard collinear PDF multiplied by a $k_T$ dependent factor taken to be a gaussian. Only the DGLAP evolution of the collinear PDF was taken into account, with the transverse momentum distribution being scale independent. While this approach is applicable when the experimental data spans a limited range of low $Q^2$ values, in general, one must take into account QCD evolution of TMD PDFs and FF's.  A proper treatment of TMD factorisation, containing well defined TMDs including their evolution properties can be found in Ref.~\cite{Collins:2011book}. Furthermore, the evolution of all leading-twist (un-)polarised TMDPDFs has been shown to be universal~\cite{Echevarria:2012pw, Echevarria:2014rua, Echevarria:2015uaa}. TMD evolution equations have  been  explicitly worked out for unpolarised TMD PDFs and FF's~\cite{Aybat:2011zv} and also for the Sivers and Collins functions~\cite{Kang:2014zza, Aybat:2011ta, Aybat:2011ge, Anselmino:2012aa}.

Although the quark Sivers functions have been studied widely, there is not much information available about the gluon Sivers function (GSF).  The first phenomenological fits of the GSF, in a GPM framework, have been obtained by D'Alesio, Murgia and Pisano ~\cite{D'Alesio:2015uta}. This was done by fitting to midrapidity data on the transverse single spin asymmetry in $pp^\uparrow\rightarrow\pi^0+X$ measured by the PHENIX collaboration. For this purpose, two different parametrisations of the quark Sivers functions (QSFs) \cite{Anselmino:2005ea,Anselmino:2008sga} ---  as extracted from polarised SIDIS processes, $lp^\uparrow\rightarrow\l'+h+X$ --- were used, leading to two different fits of the GSF. Only the DGLAP evolution of the collinear densities that occur in the TMDs was taken into account, as is standard in many low--$Q^{2}$ GPM studies. In this work, we will refer to these fits as the DMP fits. In obtaining the DMP fits, quark Sivers functions extracted using SIDIS processes, were used to constrain the gluon Sivers function in the GPM framework. In view of the earlier discussion it is then clear that one should think of processes which can probe the GSF `directly'. An example is open and closed heavy flavour production.
 
 In our previous work, we had discussed the possibility of using $J/\psi $ production in scattering of low virtuality electrons off a polarized proton to this purpose and probe the GSF {\it directly} \cite{Godbole:2013bca, Godbole:2014tha}. Therein, we had presented expected values of the asymmetry in the electroproduction of charmonium using the colour evaporation model. We had compared the expected asymmetries using DGLAP evolved TMDPDFs with those obtained using the full TMD evolution of unpolarized TMDPDFs and the GSF. Since no fits of the gluon Sivers function were available at the time, for the purpose of calculating expected values of the asymmetry, we had parametrized  the gluon Sivers function in terms of the quark Sivers functions using models by Boer and Vogelsang~\cite{Boer-PRD69(2004)094025} which we will call BV models in the current work. In these models, the collinear part of the gluon Sivers function is modelled on the collinear part of the quark Sivers functions, for which, phenomenological fits are available. 

In the present  work, we consider open heavy flavour production and perform a phenomenological study of SSA in $D$-meson production in the process  $p p^\uparrow \rightarrow D^0+X $. This was first studied in a TMD-GPM approach in Ref. \cite{Anselmino:2004nk}. In this study, two extreme values of the gluon Sivers funcion were considered i.e., zero and maximal, with only the DGLAP evolution of the collinear factorized component of the TMDs taken into account. The work showed that the measurement of asymmetry at RHIC for this process can give a direct indication of a non-zero gluon ``effective" Sivers function.  SSA in this process was also studied in the twist-3 collinear factorisation approach \cite{Kang:2008ih}.  Unpolarised Charmonium and Bottomonium scattering cross-sections within the TMD formalism and the effect of TMD evolution on these has been studied in Ref. \cite{Mukherjee:2015smo}.

Here we calculate the expected asymmetries using the two DMP fits. Furthermore, we explore the impact of TMD evolution on the asymmetry predictions. Since the DMP fits use only DGLAP evolution, one cannot achieve this exploration using the DMP fits. We therefore study the effect of TMD evolution using the extreme case of a GSF saturating the positivity bound, following Ref. \cite{Anselmino:2004nk}. We then do the same with the BV models of the GSF.

We find that the DMP fits provide small but not insignificant asymmetries that could be measurable in PHENIX. We also find that the inclusion of TMD evolution causes a significant reduction in the asymmetry predictions. This is in accordance with our previous results on $J/\psi$ electroproduction. We would also like to note that the quark contribution to the asymmetry in the kinematic regions we look at is quite small. This is mainly due to the fact that we focus on $D$-meson production in kinematic regions which are highly gluon dominated. 

In section II, we give the expresssions required to construct the asymmetry. In section III, we give the details of the DMP fits used. Section IV, contains a brief discussion of TMD evolution and the parametrizations of the TMDs to be used with it. In section V, we present the asymmetry values expected for the different cases considered. This is followed by an analysis of the results in section VI. Further details on the kinematics and the calculation are presented in the Appendix.

\section{Kinematics}
We consider the single spin asymmetry,
\be
A_N=\frac{d\sigma^\uparrow-d\sigma^\downarrow}{d\sigma^\uparrow+d\sigma^\downarrow}
\ee
for  $p^\uparrow p\rightarrow D^0+X$ at RHIC energy, $\sqrt{s} = 200\text{ GeV}$. The $D$-mesons are produced by fragmentation processes from $c$  quarks which in turn are produced at leading order by $q\bar{q}$ annihilation or $gg$ fusion processes, the former being a purely $s$-channel process. The gluon fusion process thus dominates $D$-meson production. Since both the gluon fusion and $q\bar q$ annihilation result in unpolarized final state partons, there can be no contribution to the SSA from the Collins effect. Furthermore, it has been verified that no other spin related TMDs apart from the Sivers function contribute to SSA in this process \cite{Anselmino:2004nk}. 

Following Ref. \cite{Anselmino:2004nk}, we write for the numerator of the asymmetry,
\bea
d\sigma ^\uparrow - d\sigma ^\downarrow &=& 
\frac{E_D \, d\sigma^{\pup p \to DX}} {d^{3} \bfp_D} -
\frac{E_D \, d\sigma^{\pdown p \to DX}} {d^{3} \bfp_D}  
\label{final-ssa} \\
&& \hspace*{-3.0cm} = \>
\int dx_a \, dx_b  \, dz \, d^2 \mathbf{k}_{\perp a} \, d^2 \mathbf{k}_{\perp b} \, 
d^3 \mathbf{k}_{D} \, 
\delta (\mathbf{k}_{D} \cdot \hat{\bfp}_c) \, 
\delta (\hat s +\hat t +\hat u - 2m_c^2) \> 
{\mathcal C}(x_a,x_b,z,\mathbf{k}_D) \nonumber \\
&& \hspace*{-3.0cm} \times \, \Biggl\{ \sum_q
\left[\Delta ^N f_{q/\pup}(x_a,\mathbf{k}_{\perp a}) \>  f_{\bar q/p}(x_b, 
\mathbf{k}_{\perp b}) \>
\frac{d \hat{\sigma}^{q \bar q \to c \bar c}}
{d\hat t}(x_a, x_b, \mathbf{k}_{\perp a}, \mathbf{k}_{\perp b}, \mathbf{k}_D) \>
 D_{D/c}(z,\mathbf{k}_D) \right]  \nonumber \\
&& \hspace*{-3.0cm} +
\left[ \Delta ^N f_{g/\pup}(x_a,\mathbf{k}_{\perp a}) \>  f_{g/p}(x_b, 
\mathbf{k}_{\perp b}) \>
\frac{d \hat{\sigma}^{gg \to c \bar c}}
{d\hat t}(x_a, x_b, \mathbf{k}_{\perp a}, \mathbf{k}_{\perp b}, \mathbf{k}_D) \>
 D_{D/c}(z,\mathbf{k}_D) \right] \Biggr\} \>, \nonumber 
\eea
where $q=u,d,s,\bar{u},\bar{d},\bar{s}$ and the denominator is given by,
 \bea
d\sigma ^\uparrow + d\sigma ^\downarrow &=& 
\frac{E_D \, d\sigma^{\pup p \to DX}} {d^{3} \bfp_D} +
\frac{E_D \, d\sigma^{\pdown p \to DX}} {d^{3} \bfp_D}
= 2 \, \frac{E_D \, d\sigma^{pp \to DX}} {d^{3} \bfp_D} 
\label{final-unp} \\ 
&& \hspace*{-2.5cm} = 2
\int dx_a \, dx_b \, dz\, d^2 \mathbf{k}_{\perp a} \, d^2 \mathbf{k}_{\perp b} \, 
d^3 \mathbf{k}_D \, 
\delta (\mathbf{k}_D \cdot \hat{\bfp}_c) \, 
\delta (\hat s +\hat t +\hat u - 2m_c^2) \, {\mathcal C}(x_a,x_b,z,\mathbf{k}_D)
\nonumber \\
&&  \hspace*{-2.5cm} \times \, \Biggl\{ \sum_q
\left[ \hat f_{q/p}(x_a,\mathbf{k}_{\perp a}) \> 
\hat f_{\bar q/p}(x_b, \mathbf{k}_{\perp b}) \>
\frac{d \hat{\sigma}^{q \bar q \to c \bar c}}
{d\hat t}(x_a, x_b, \mathbf{k}_{\perp a}, \mathbf{k}_{\perp b},  \mathbf{k}_D) \>
\hat D_{D/c}(z,\mathbf{k}_D) \right] \nonumber \\
&& \hspace*{-2.5cm} +
\left[ \hat f_{g/p}(x_a,\mathbf{k}_{\perp a}) \> \hat f_{g/p}(x_b, \mathbf{k}_{\perp b}) \>
\frac{d \hat{\sigma}^{gg \to c \bar c}}
{d\hat t}(x_a, x_b, \mathbf{k}_{\perp a}, \mathbf{k}_{\perp b}, \mathbf{k}_D) \>
\hat D_{D/c}(z,\mathbf{k}_D) \right] \Biggr\} \>. \nonumber 
\eea
In the above expressions, $x_{a,b}$ are the light-cone momentum fractions of the incoming partons along the parent proton direction, $z=p_D^+/p_c^+$ is the light-cone momentum fraction of the $D$-meson along the fragmenting parton direction, $\mathbf{k}_{\perp a,b}$ are the intrinsic transverse momenta of the incoming partons with respect to parent proton direction, $\mathbf{k}_D$ is the transverse momentum with which the meson fragments from the heavy quark, $\hat{\bfp}_c$ is the unit vector along the heavy quark direction, $m_c$ is the heavy quark mass, and $\hat s$, $\hat t$ and $\hat u$ are the partonic mandelstam variables defined in the standard way.

The expressions $\Delta^N f_{i/\pup}(x,\mathbf{k}_\perp)$ and $f_{i/p}(x,\mathbf{k}_\perp)$ are the Sivers function and the unpolarised TMDPDF for the flavour $i$ respectively. $D_{D/c}(z,\mathbf{k}_D)$ is the transverse momentum dependent fragmentation function (TMDFF). Their functional forms are given in Sections 3 and 4. 

The Sivers distribution $\Delta^Nf_{i/p^\uparrow}$, for an unpolarised parton $i$ inside a transversely polarized proton is defined as,
\bea
f_{i/p^\uparrow}(x,\mathbf{k}_\perp,\mathbf{S};Q)&=&f_{i/p}(x,k_\perp;Q)-f^{\perp i}_{1 T}(x,k_\perp;Q)\frac{\epsilon_{ab}k^a_\perp S^b}{M_p}
\\
&=&f_{i/p}(x,k_\perp;Q)+\frac{1}{2}\Delta^N f_{i/p^\uparrow}(x,k_\perp;Q)\frac{\epsilon_{ab}k_\perp^a S^b}{k_\perp}.
\eea

$C(x_a,x_b,z, \mathbf{k}_D)$ contains the flux and jacobian factors for the transformation from a partonic to a mesonic phase space. It is given by,
\be
{\mathcal C} = \frac{\hat s}{\pi z^2}\,\frac{\hat s}{x_a x_b s}\,
\frac{ \left( E_D+\sqrt{\mathbf{p}_D^2 - \mathbf{k}_{\perp D}^2} \right)^2}
{4(\mathbf{p}_D^2 - \mathbf{k}_{\perp D}^2)} \,
\left[1- \frac{z^2 m_c^2}
{ \left( E_D+\sqrt{\mathbf{p}_D^2 - \mathbf{k}_{\perp D}^2} \right)^2}\right]^2
\,.
\ee

The partonic cross-sections are given by,
\bea
\frac{d\hat\sigma ^{q\bar q\to Q\bar Q}}{d\hat t} &=& 
\frac {\pi \alpha_s^2}{\hat s^2} \, \frac {2}{9}
\left(2\tau_1^2 + 2\tau_2^2 + \chi \right) \>,
\label{qqunp} \\
\frac{d\hat\sigma ^{gg\to Q\bar Q}}{d\hat t} &=& 
\frac {\pi \alpha_s^2}{\hat s^2} \, \frac {1}{8}\,
\left(\frac{4}{3\tau_1\tau_2}-3\right)
\left(\tau_1^2 + \tau_2^2 + \chi -\frac{\chi^2}{4\tau_1\tau_2} \right) \>,
\label{ggunp} 
\eea
where $\tau _{1,2}$ and $\chi$ 
are dimensionless quantities given by, 
\be
\tau_1 = \frac{m_Q^2-\hat t}{\hat s}, \,\,\,
\tau_2 = \frac{m_Q^2-\hat u}{\hat s},\,\,\,
\chi = \frac{4m_Q^2}{\hat s}\,.
\label{tau}
\ee

We determine the value of $z$ by solving the on-shell condition given by,
\be
\hat s+\hat t+\hat u=2m_c^2.
\ee
This is reflected in the $\delta$ function $\delta(\hat s+\hat t+\hat u-2m_c^2)$ in Eqs. 2 and 3
to fix the value of $z$. Here, the mandelstam variables $\hat s$, $\hat t$ and $\hat u$ are defined in the usual way in terms of the partonic variables as,
\bea
\hat s = (P_a+P_b)^2&=x_ax_bs\left[1-2\frac{k_{\perp a}k_{\perp b}}{x_ax_bs}\cos(\phi_a-\phi_b)+\frac{k_{\perp a}^2k_{\perp b}^2}{x_a^2x_b^2s^2}\right]
\\ \nonumber
\hat t = (P_a-P_c)^2&
\\ \nonumber
\hat u = (P_b-P_c)^2&
\eea
We have not given the expressions for $\hat t$ and $\hat u$ as they are lengthy and complicated.

For the case of a massless parton fragmenting into mesons, it is possible to express $\hat t$ and $\hat u$ as $T/z$ and $U/z$ respectively \cite{D'Alesio:2004up}. This allows the value of $z$ to be uniquely determined from the on-shell condition. However, in the case of heavy partons, the expressions for $\hat t$ and $\hat u$ depend on $z$ in a much more complicated manner and the on-shell delta function cannot be written in the simple form of Eq. (10) above. Keeping the $z$ dependencies explicit, the on-shell condition takes the form,
\be
\label{onshellai}
\frac{a_1}{z}+a_2 z+a_3\sqrt{-a_4+\frac{a_5(1+\frac{a_4 z^2}{a_5})^2}{4z^2}}+a_6=0
\ee
where the factors $a_i$ do not depend on $z$. This is a quartic equation in $z$ and has four solutions given in terms of $a_i$. We find that only one of these solutions gives physical values of $z$. The expressions for  $a_i$ and the solution for $z$ in terms of them are given in the appendix.

\section{Parametrization of the TMDs}
\label{param-tmds}
For the predictions with the DMP fits, we use the same functional forms for the TMDs as in Ref.~\cite{D'Alesio:2015uta}. For the unpolarised TMDPDF, we use the standard factorised gaussian form given by:
\be
f_{i/p}(x,k_\perp;Q)=f_{i/p}(x,Q)\frac{1}{\pi\langle k_\perp^2\rangle}e^{-k_\perp^2/\langle k_\perp^2\rangle}
\ee
with $\langle k_\perp^2\rangle=0.25\text{ GeV}^2$ and  $i=q,g$.

%For the Sivers function, we use the parametrization given in Ref. \cite{D'Alesio:2015uta},
The Sivers function is parametrized as,
\be
\Delta^N f_{i/p^\uparrow}(x,k_\perp;Q)=2\mathcal{N}_{i}(x)f_{i/p}(x,Q)h(k_\perp)\frac{e^{-k^2_\perp/\langle k_\perp^2\rangle}}{\pi \langle k_\perp^2\rangle}
\ee
with, 
\be
\mathcal{N}_i(x)=N_i x^{\alpha_i}(1-x)^{\beta_i}\frac{(\alpha_i+\beta_i)^{\alpha_i+\beta_i}}{\alpha_i^{\alpha_i} \beta_i^{\beta_i}}
\label{Nx}
\ee
and
\be
h(k_\perp)=\sqrt{2e}\frac{k_\perp}{M_1}e^{-k_\perp^2/M_1^2}
\ee
where $M_1$ is a parameter which is determined by fits to data on SSAs and $e$ is Euler's number.

The $k_\perp$-dependent part of the Sivers function can be expressed in terms of another parameter $\rho$ as follows:
\be
h(k_\perp)\frac{e^{-k^2_\perp/\langle k_\perp^2\rangle}}{\pi \langle k_\perp^2\rangle}=\frac{\sqrt{2e}}{\pi}\sqrt{\frac{1-\rho}{\rho}}k_\perp \frac{e^{-k^2_\perp/\rho\langle k^2_\perp\rangle}}{\langle k^2_\perp\rangle^{3/2}}
\ee
where,
\be
\rho=\frac{M_1^2}{\langle k^2_\perp\rangle+M_1^2}.
\label{rho}
\ee
%We give this particular form in Eq.~18 since the DMP fits were extracted using this parametrization.
We give this particular form in Eq.~18 since the DMP fits are given in terms of $\rho$.

The two fits of the GSF that we use, which have been referred to as SIDIS1 and SIDIS2~\cite{D'Alesio:2015uta}, have been obtained by fitting to the PHENIX data on pion production in the mid-rapidity region at RHIC in the process $pp^\uparrow\rightarrow\pi^0+X$. The two differ in the parametrisations of the quark Sivers functions (QSFs) that have been used. SIDIS1 is obtained with a parametrisation of the QSFs~\cite{Anselmino:2005ea} which contains only the $u$ and $d$ flavours, with the input being data on pion production from the HERMES experiment and data on positive hadron production from the COMPASS experiment. SIDIS2 is obtained with a parametrisation of the QSFs~\cite{Anselmino:2008sga} where flavour segregated data on pion and kaon production is used and hence all three light flavours are included. Further details on the differences between the two fits can be found in Ref.~\cite{D'Alesio:2015uta}. The values of the parameters of the two fits are given in Table I.
\begin{table}[H]
\centering
\begin{tabular}{|l|l|l|l|l|l|l|}
\hline
SIDIS1 & \multicolumn{2}{l|}{$N_g=0.65$} & $\alpha_g=2.8$ & $\beta_g=2.8$ & $\rho=0.687$ & \multirow{2}{*}{$\langle k^2_\perp\rangle=0.25 GeV^2$} \\ \cline{1-6}
SIDIS2 & \multicolumn{2}{l|}{$N_g=0.05$} & $\alpha_g=0.8$ & $\beta_g=1.4$ & $\rho=0.576$ &                                                        \\ \cline{1-7}
\end{tabular}
\caption{Parameters of the DMP fits.}
\label{SIDIS-gluon-fits}
\end{table}
We give predictions for the gluon Sivers asymmetry using these two Sivers functions.

\section{TMD Evolution}
\label{tmd-evolution}
Below we give a brief outline of the evolution of the transverse momentum dependent functions as given in Ref. \cite{Anselmino:2012aa}. This was referred to as TMD-e1 in our previous work \cite{Godbole:2014tha}. TMDs can be written in coordinate space (called $b$-space) as a fourier transform given by,
\be
F(x,b;Q)=\int d^2k_\perp e^{-i\vec k_\perp .\vec b}F(x,k_\perp;Q).
\ee
Since the $Q^2$ evolution is more naturally described in $b$-space, we choose to work with $b$-space TMDs. The $Q^2$ evolution of $b$-space TMDs is given by,
\be
F(x,b,Q_f)=F(x,b,Q_i) R_\text{pert}(Q_f,Q_i,b) R_\text{NP}(Q_f,Q_i,b)
\ee
where $R_\text{pert}$ is the perturbative part of the evolution kernel and $R_\text{NP}$ is the nonperturbative part.

The perturbative part is given by,
\be
R_\text{pert} (Q_f,Q_i,b) \equiv \exp\left\{\ln\int^{\mu_b}_{Q_i}\frac{d\mu}{\mu}\gamma_K(\mu) +\int^{Q_f}_{Q_i}\frac{d\mu}{\mu}\gamma_F\left(\mu,\frac{Q^2}{\mu^2}\right)\right\}
\ee
The various quantities appearing in above equations are as follows: $\gamma_K$ and $\gamma_F$ are anomalous dimensions which are different for quarks and gluons; $\mu_b=2e^{-\gamma_E}/b_*(b_T)$, where $b_*(b_T)=b_T/\sqrt{1+b_T^2/b_\text{max}^2}$ is the standard prescription used to stitch together the perturbative and nonperturbative parts of the kernel, and $\gamma_E$ is the Euler-Mascheroni constant;
the anomalous dimensions $\gamma_K$ and $\gamma_F$ at order $\mathcal{O}(\alpha_s)$ are  \cite{Echevarria:2015uaa},

\be
\gamma_F(\mu;\frac{Q^2}{\mu^2})=\alpha_s(\mu)\frac{C_F}{\pi}\left(\frac{3}{2}-\ln\frac{Q^2}{\mu^2}\right), \hspace{0.5cm} \gamma_K(\mu)=\alpha_s(\mu)\frac{2C_F}{\pi}
\ee
for quarks and,
\be
\gamma_F(\mu;\frac{Q^2}{\mu^2})=\alpha_s(\mu)\left(-\frac{C_A}{\pi}\ln\frac{Q^2}{\mu^2}-\frac{1}{2}\left(\frac{11}{3}C_A-\frac{2}{3}N_f\right)\right), 
\hspace{0.5cm}
\gamma_K(\mu)=\alpha_s(\mu)\frac{2C_A}{\pi} 
\label{anomalous-dimensions}
\ee
for gluons. Finally, the nonperturbative exponential part, the Sudakov factor is given by,
\be
R_\text{NP}\equiv\exp\left\{-\frac{1}{2}g_2b^2_T\ln\frac{Q}{Q_i}\right\}
\ee
Here, following Ref. \cite{Anselmino:2012aa}, we use $g_2=0.68$, corresponding to a $b_\text{max}=0.5\text{ GeV}^{-1}$.

\subsection{Parametrization of $b$-space TMDs at initial scale}
Just as in the DGLAP case, the unpolarised TMDPDF at the initial scale is chosen to be a gaussian. The following exponential form for the corresponding $b$-space density,
\be
f_{i/p}(x,b_T;Q_0)=f_{i/p}(x,Q_0)\exp\{-\langle k_\perp^2\rangle b_T^2/4\},
\ee
after fourier transforming, gives the commonly used gaussian distribution in the transverse momentum space:
\be
f_{i/p}(x,k_\perp;Q_0)=f_{i/p}(x,Q_0)\frac{1}{\pi \langle k_\perp^2\rangle}\exp\{-k_\perp^2/\langle k_\perp^2\rangle\}
\ee
where $i=q,g$ and the initial scale $Q_0=1$ GeV.

The evolution of the Sivers function is obtained through its first derivative in $b$-space. This is parametrized at the initial scale as,
\be
f'^{\perp i}_{1T}(x,b_T;Q_0)=-\frac{\rho\langle k_\perp^2\rangle}{2}f^{\perp i}_{1T}(x;Q_0)b_T \exp\{-\frac{\rho\langle k_\perp^2\rangle}{4} b_T^2\}
\ee
where $\rho$ is of the form given in Eq.~\ref{rho}
and
\be
f^{\perp i}_{1T}(x;Q_0)=\frac{M_p}{2M_1}\sqrt{2e}\Delta^Nf_{i/p^\uparrow}(x,Q_0)\rho,
\ee
where $M_p$ is the mass of the proton and $\Delta^Nf_{i/p^\uparrow}(x,Q_0)$ is the $x$-dependent part of the Sivers function at the initial scale. It is written as,
\be
\Delta^Nf_{i/p^\uparrow}(x,Q_0)=2\mathcal{N}_{i}(x)f_{i/p}(x,Q_0).
\ee
Here, $\mathcal{N}_{i}(x)$ has the same form as in Eq \ref{Nx}.

The above form ensures that the $k_\perp$-space Sivers function at the initial scale has same form as in the DGLAP case:
\be
\Delta^Nf_{i/p^\uparrow}(x,k_\perp;Q_0)=2\mathcal{N}_{i}(x)h(k_\perp)f_{i/p}(x,k_\perp;Q_0)
\ee
where,
\be
h(k_\perp)=\sqrt{2e}\frac{k_\perp}{M_1}e^{-k_\perp^2/M_1^2}.
\ee
Here,  $|\mathcal{N}_{i}(x)|\leq 1$ and $h(k_\perp)\leq 1$ so the Sivers function always obeys the positivity bound given by:
\be
\frac{|\Delta^Nf_{i/p^\uparrow}(x,k_\perp)|}{2f_{i/p}(x,k_\perp)}\leq 1.
\label{saturation-bound}
\ee
\subsection{Transverse momentum-space TMDs}
The expressions for the TMDs in $k_\perp$-space can be obtained by fourier transforming the $b_T$-space expressions:
\bea
f_{i/p}(x,k_\perp;Q)&=&\frac{1}{2\pi}\int^\infty_0\text{d}b_Tb_TJ_0(k_\perp b_T)f_{i/p}(x,b_T;Q)\\
f^{\perp i}_{1T}(x,k_\perp;Q)&=&\frac{-1}{2\pi k_\perp}\int^\infty_0\text{d}b_Tb_TJ_1(k_\perp b_T)f'^{\perp i}_{1T}(x,b_T;Q)
\eea
The above expression for the Sivers function is related to $\Delta^N f_{i/p^\uparrow}$ through Eq. 4, as follows:
\be
\Delta^N f_{q(g)/p^\uparrow}(x,k_\perp)=-\frac{2k_\perp}{M_p}f^{\perp q(g)}_{1T}(x,k_\perp).
\ee
Unlike in the DGLAP case, we don't have any fits of the GSF to data obtained using TMD evolution. Hence to illustrate the suitability of the probe we consider three cases:
\begin{enumerate}
\item A maximal gluon Sivers function obtained by saturating the bound given in Eq. \ref{saturation-bound}. This is obtained by setting $\mathcal{N}_g(x)$ to 1. We refer to this as the ``saturated" GSF.
\item Sivers function with $\mathcal{N}_g(x)=(\mathcal{N}_u(x)+\mathcal{N}_d(x))/2$ \hspace{0.5cm} (BV (A))
\item Sivers function with $\mathcal{N}_g(x)=\mathcal{N}_d(x)$  \hspace{0.5cm} (BV (B))
\end{enumerate}
We will compare the predictions obtained with DGLAP and TMD evolution for each of these cases. The last two are models proposed by Boer and Vogelsang \cite{Boer-PRD69(2004)094025} in which the $x$-dependent part of the gluon Sivers function is modelled on quark Sivers functions.  $\mathcal{N}_u(x)$ and $\mathcal{N}_d(x)$ are of the form given in Eq.~\ref{Nx} with their parameters being given in Ref. \cite{Anselmino:2012aa} for the case of TMD evolution. For the case of DGLAP evolution, we choose to use the parameters given in Ref. \cite{Anselmino:2005ea} which were used to obtain the SIDIS1 GSF fit. The BV parametrizations give a GSF with an opposite sign relative to the saturated GSF and DMP fits presented in Section III.

In obtaining our predictions we used the same gaussian width $\langle k^2_\perp\rangle=0.25\text{ GeV}^2$ for the gluon as was used for the quarks in Ref.~\cite{Anselmino:2012aa}. The values of $M_1$ and the parameters for $\mathcal{N}_u(x)$ and $\mathcal{N}_d(x)$, obtained by fits to data, were also taken from the same and are given in the appendix.

\section{Results}
In this section, we present cross-section and asymmetry predictions obtained with and without TMD evolution.   We first present results for RHIC centre of mass energy $\sqrt{s}=200\text{ GeV}$. Then, keeping planned experiments in mind, we also consider the possibilities of probing the GSF at AFTER@LHC ($\sqrt{s} = $115 GeV)\cite{Rakotozafindrabe:2013au, Lansberg:2016urh} and at a future RHIC run ($\sqrt{s} = $ 500 GeV)\cite{Aschenauer:2015eha}. For all c.o.m energies, we consider the range $-0.7\leq x_F\leq0.7$ at a fixed meson $P_T=1.5 \text{ GeV}$, and the range $0.5\text{ GeV}\leq P_T\leq3.5\text{ GeV}$ at fixed meson pseudorapidity values $\eta=2.0,3.8$.  We use the GRV98-LO pdf set for the unpolarized parton distributions and for the collinear part of the charm quark fragmentation functions, we use those given by Cacciari {\it et  al} \cite{Cacciari:1996wr}.

\begin{figure}[h]
\subfigure[]{\includegraphics[width=0.49\linewidth]{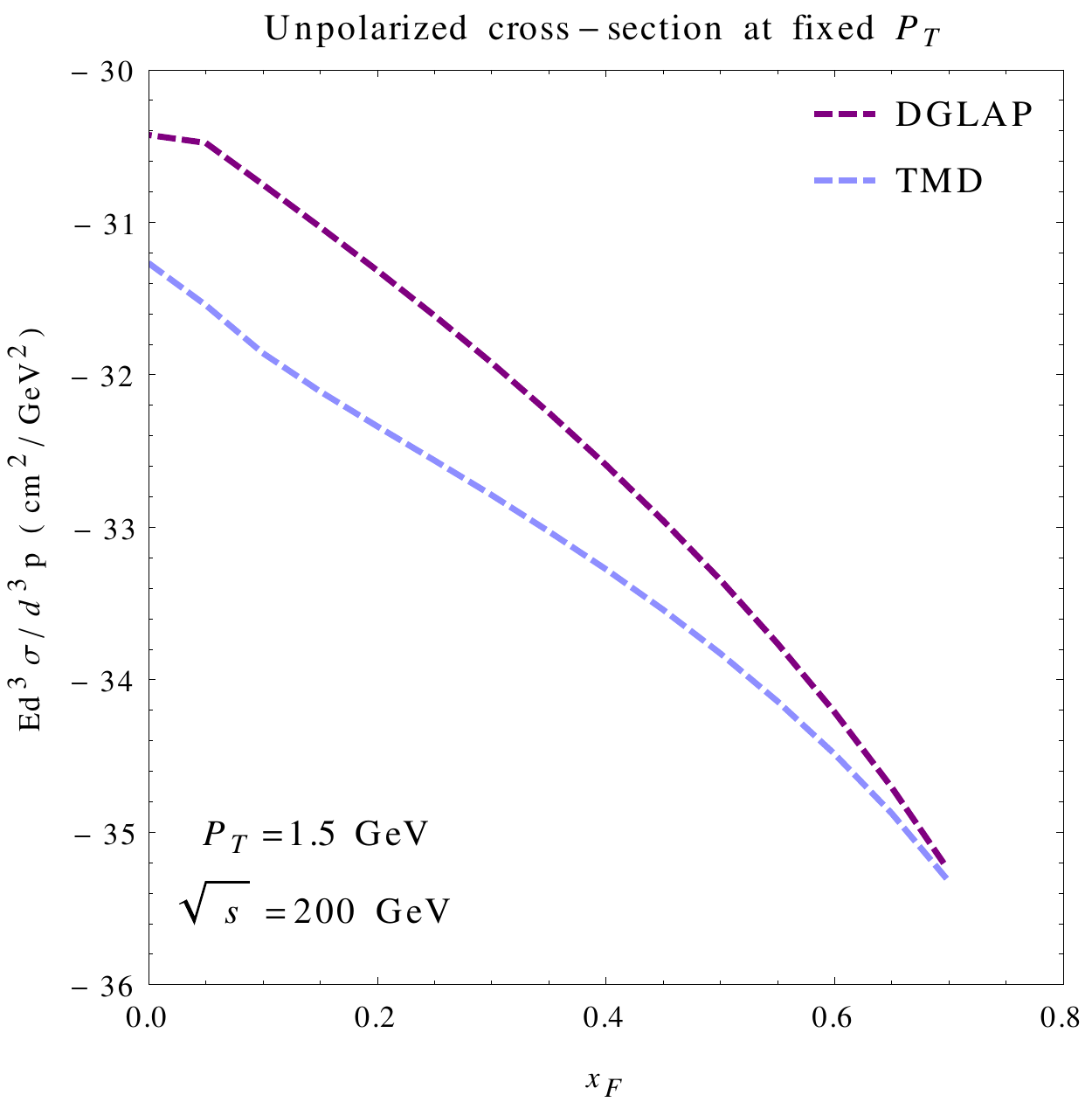}}
\subfigure[]{\includegraphics[width=0.48\linewidth]{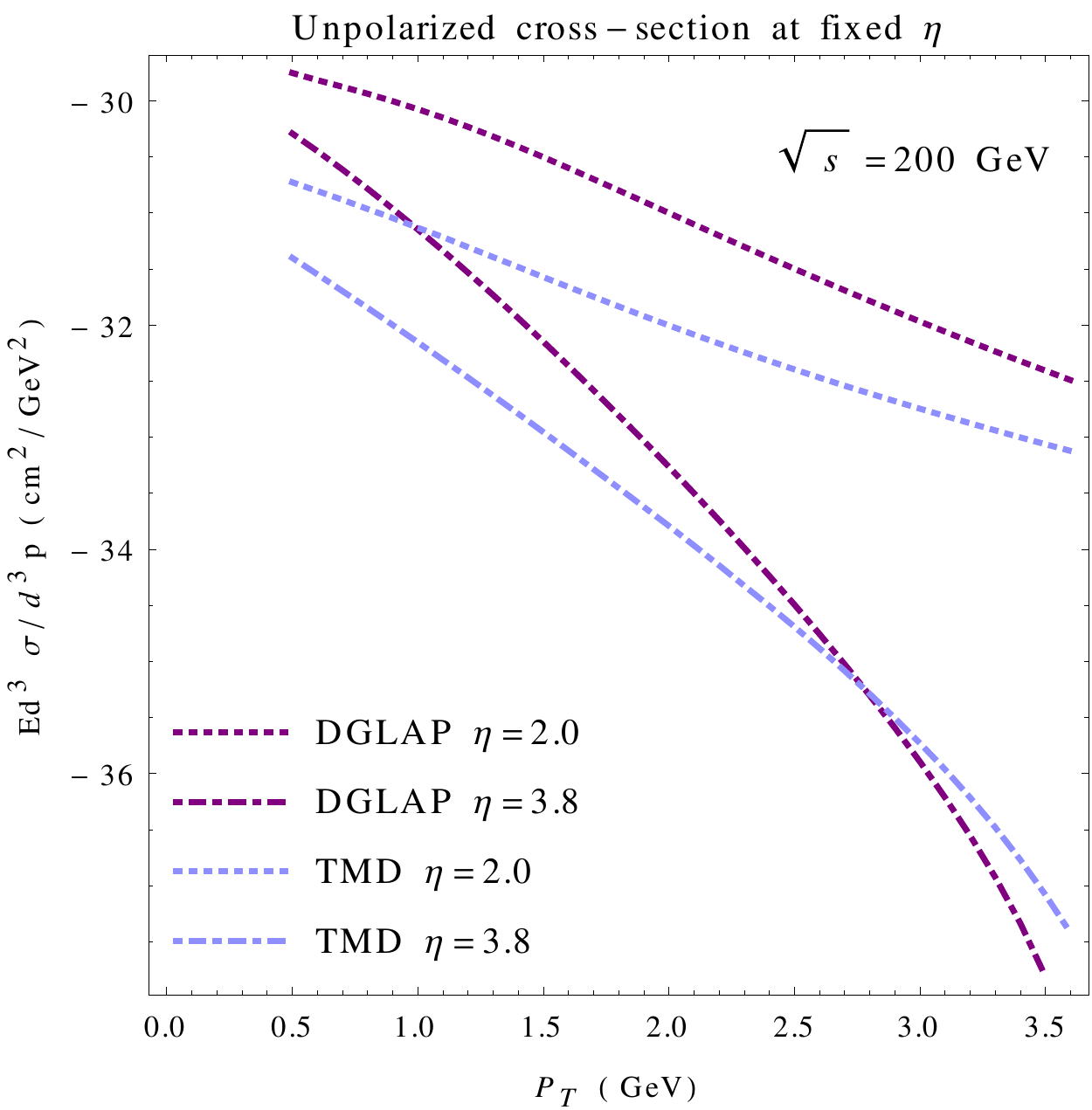}}
\caption{Unpolarized cross-section: Panel (a) (left) shows the numbers at fixed $P_T=1.5$ GeV and (b) (right) shows the numbers at fixed pseudorapidity values $\eta=2.0,3.8$. The red dashed line is for the results obtained using DGLAP evolution with $\langle k_\perp^2\rangle=0.25\text{ GeV}^2$ (c.f Section III) and the blue dotted line denotes the results obtained using TMD evolution (c.f Section IV).}
% Fig. 1: Unpolarized cs
\label{unpol}
\end{figure}

We begin with the results for the unpolarized cross-section presented in Fig. \ref{unpol}.
As can be seen from the above plot, the values obtained for the unpolarized cross-section using DGLAP and TMD evolved schemes, differ from each other in all $P_T$ (for fixed $\eta$) and low-to-moderate $x_F$ (for fixed $P_T$) regions by almost an order of magnitude. The magnitude of the cross-section decides the accuracy of the asymmetry measurements. This stresses the need for a proper measurement of $D$-meson production in these kinematics. This process may also serve as a good probe with which to fit the parameters of the unpolarized gluon TMDPDF.  
The cross-section at $\eta=3.8$ is found to be much smaller than that at $\eta=2.0$. This is because  larger $x$ values contribute to the production at a given $P_T$ for $\eta=3.8$, compared to those for $\eta=2.0$. Further, with increasing $P_T$, the cross-section predictions decrease faster for $\eta=3.8$ as the $x$-values that contribute increase further than for $\eta=2.0$.

For the asymmetry results obtained without TMD evolution, i.e., the DGLAP case (c.f Section III), we primarily show the predictions obtained with the DMP fits \cite{D'Alesio:2015uta}. Note that these are the only available GSF parametrisations obtained using data. 

For the case of TMD evolution (c.f. Section IV), there are no available fits of the gluon Sivers function using data. Hence we present results for following two cases - 1) Using a maximal, saturated GSF and 2) using the BV (A) and (B) models of the GSF.  In each case we compare the predictions with corresponding ones obtained using just DGLAP evolution.

\begin{figure}[h]
\subfigure[]{\includegraphics[width=0.49\linewidth]{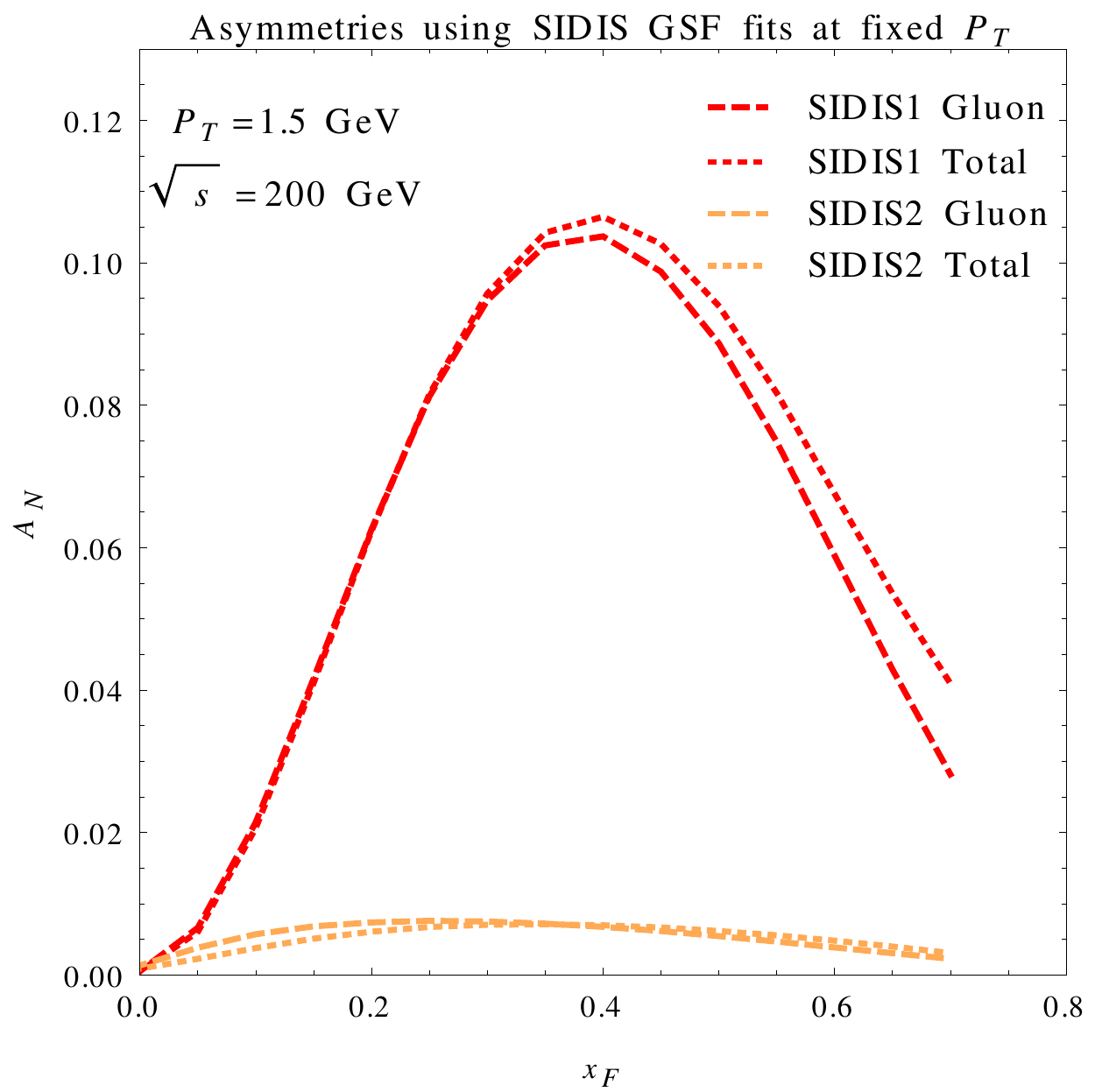}}
\subfigure[]{\includegraphics[width=0.49\linewidth]{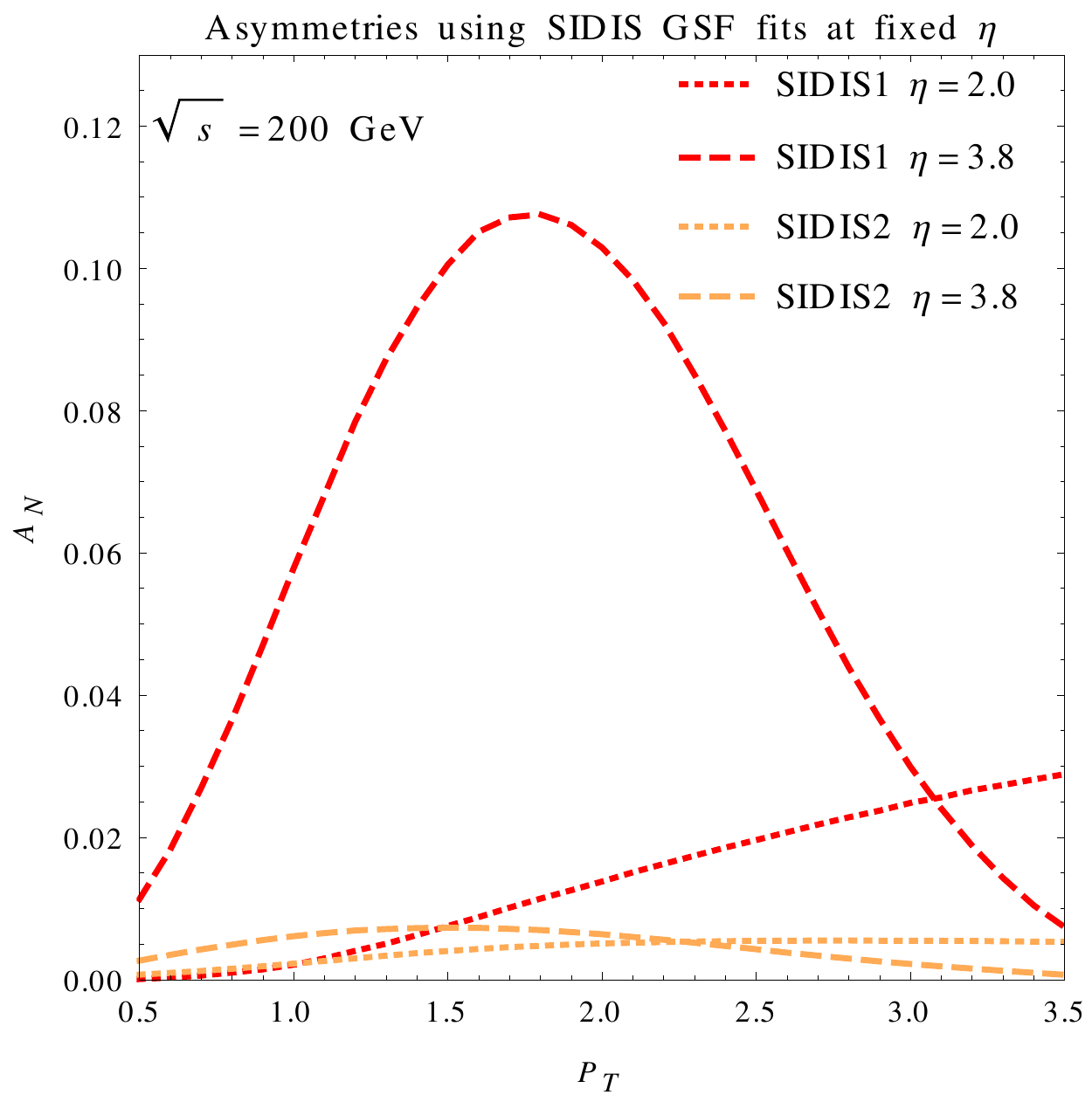}}
\caption{Asymmetry predictions using the DMP fits: Panel (a) (left) shows results fixed $P_T=1.5$ GeV and panel (b) (right) shows results at fixed $\eta=2.0,3.8$. Predictions using the SIDIS1 GSF are in red and those using the SIDIS2 GSF are in orange. On the left panel, we show results obtained when the quark contribution is also included. As can be seen, it is relatively small.}
\label{as_sidis_gluon}
% Fig. 2: SIDIS Gluon Asymmetries
\end{figure}
In Fig. \ref{as_sidis_gluon}, we show the gluon contribution to the asymmetry obtained using the DMP gluon fits. In Fig. 2 (a), for the results at fixed $P_T$, we also show the contribution to the asymmetry from quark Sivers functions. As expected from the numbers shown in Table I, the SIDIS1 asymmetries are substantially larger than the SIDIS2 asymmetries, with a peak value of 11\% as opposed to roughly 0.5\%. The SIDIS1 asymmetries at fixed $P_T=1.5$ and fixed $\eta=3.8$ are actually of the same order of magnitude as the maximum asymmetries obtained with a saturated GSF (Fig. 3). For $\eta=2.0$, the asymmetries from both fits rise with increasing $P_T$ as opposed to the case for $\eta=3.8$, where they peak at intermediate values in the $P_T$ range considered. While the SIDIS1 estimate at $\eta=2.0$ (peaks at 3\%) is in general lower than that at $\eta=3.8$ (peaks at 11\%) in the $P_T$ range considered, it must be kept in mind the the cross-sections (Fig. \ref{unpol} (b)) at $\eta=2.0$ are much higher. This may possibly make it easier to measure a smaller asymmetry as the statistical error on the asymmetry measurement decreases as $\frac{1}{\sqrt{N}}\sqrt{1-A^2}\approx1/\sqrt{N}$, where $N$ is the number of events. 

In general, the quark contribution to the asymmetry is much smaller than the gluon contribution and hence we do not show it. In Fig. \ref{as_sidis_gluon} (a), where we have included it, it can be seen that the gluon contribution is indeed
dominant. For all other cases (asymmetries with DMP fits at fixed $\eta$ and predictions with TMD evolved densities), the relative size of the quark contribution is even smaller, contributing at less than 5\% at peak values of the total asymmetry. In general, we find that the ratio of contributions to the asymmetry of gluons over quarks, increases with the pseudorapidity for all the Sivers functions considered. This, combined with the fact that the asymmetries themselves are larger (by about an order of magnitude in the low $P_T$ region) may make it worthwhile to measure the asymmetry at larger values of pseudorapidity even though the cross-sections themselves are much smaller in the large $\eta$ regions (cf. Fig. 1 (b)).

In Fig. 3 and 4 we compare asymmetries obtained with DGLAP evolved densities with those obtained with TMD evolved densities. We do this to demonstrate the effect of taking into account TMD evolution. Fig. 3 shows the results for the choice of a maximal saturated GSF (obtained by saturating the positivity bound in Eq. \ref{saturation-bound} for all values of $x$), and 
Fig. 4 shows the results for the BV (A) and (B) models of the GSF defined in section \ref{tmd-evolution}-B. For the DGLAP results in Fig. 4, we used the quark Sivers function parameters from the SIDIS1 fit in the BV models.  In general, a significant reduction of the asymmetry predictions is observed. For a saturated GSF (Fig. 3) the peak asymmetry with TMD evolution drops to a third of its value for DGLAP evolution. The BV models give sizeable peak values of the asymmetry in the range $1\%\lesssim|A_N|\lesssim5\%$ for TMD evolved densities and $3\%\lesssim|A_N|\lesssim9\%$ for DGLAP evolved densities, with the results obtained with TMD evolved densities always being smaller than the corresponding DGLAP results.  While the predictions from the BV models may be of similar sizes as the predictions obtained using a saturated GSF, they have an opposite sign in most kinematic regions.

\begin{figure}[h]
\subfigure[]{\includegraphics[width=0.49\linewidth]{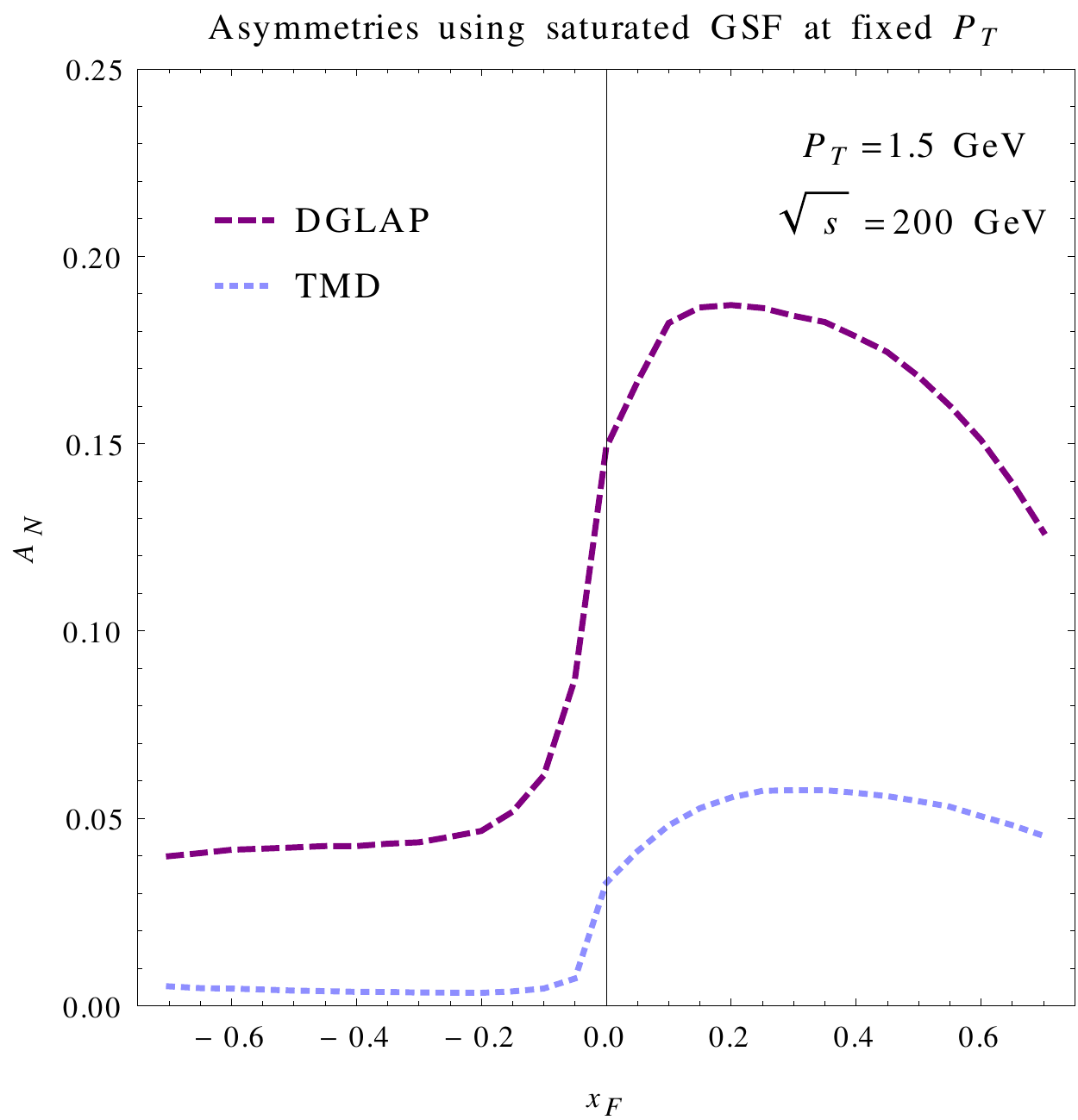}}
\hspace*{0.1cm}
\subfigure[]{\includegraphics[width=0.49\linewidth]{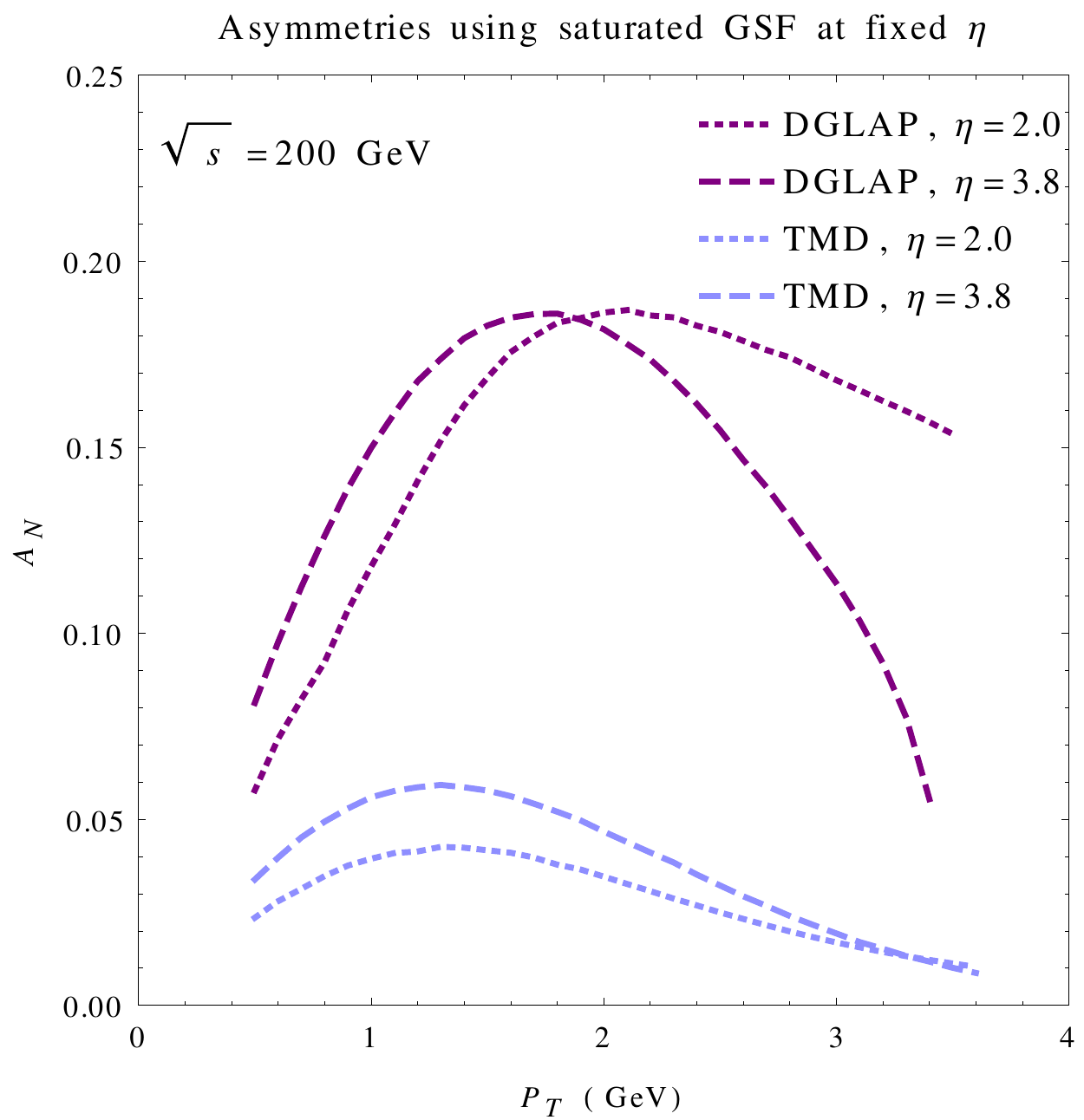}}
\caption{Asymmetry predictions using a saturated GSF evolved with DGLAP and TMD evolution: Panel (a) (left) shows results fixed $P_T=1.5$ GeV and panel (b) (right) shows results at fixed $\eta=2.0,3.8$. Results obtained with DGLAP densities are in violet and those obtained with TMD evolved densities are in blue.}
\label{comp_dglaptmd}
% Fig. 3: Saturated GSF asymmetries.
\end{figure}

\begin{figure}[h]
\subfigure[]{\includegraphics[width=0.49\linewidth]{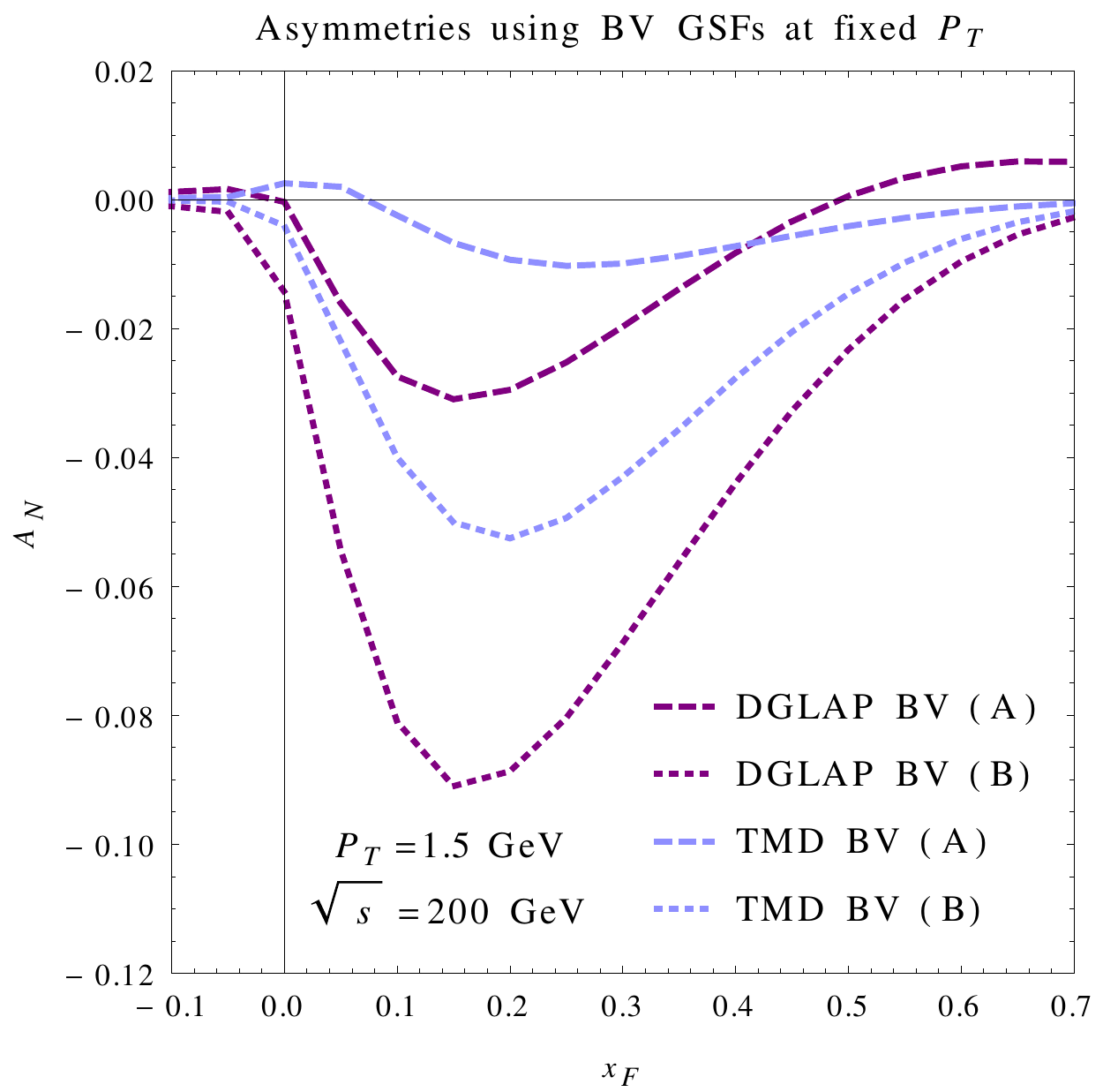}}
\hspace*{0.1cm}
\subfigure[]{\includegraphics[width=0.49\linewidth]{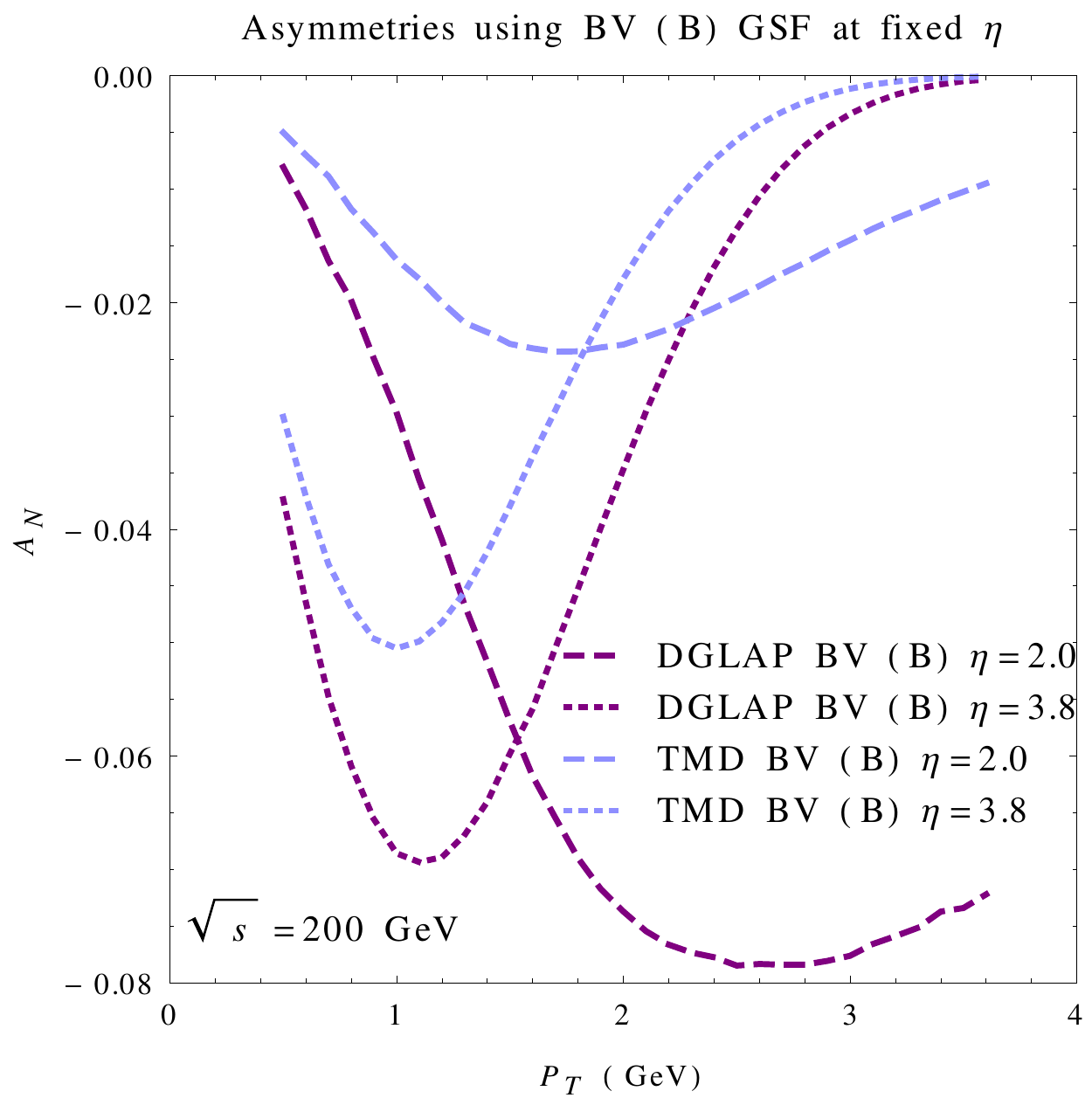}}
\caption{Asymmetry predictions obtained for the BV models of the GSF (BV (A) - dashed, BV (B) - dotted; c.f. Section IV.B) using DGLAP (in violet) and TMD evolved (in blue) densities. SIDIS1 quark Sivers function parameters were used for the DGLAP BV models. Panel (a) (left) shows results fixed $P_T=1.5$ GeV and panel (b) (right) shows results at fixed $\eta=2.0,3.8$. At fixed pseudorapidity, only results with BV (B) are shown }
\label{as_BV}
%Fig. 4: BV asymmetries using DGLAP and TMD.
\end{figure}

\subsection*{Dependence of the results on $\sqrt{s}$}
\begin{figure}[h]
\subfigure[]{\includegraphics[width=0.48\linewidth]{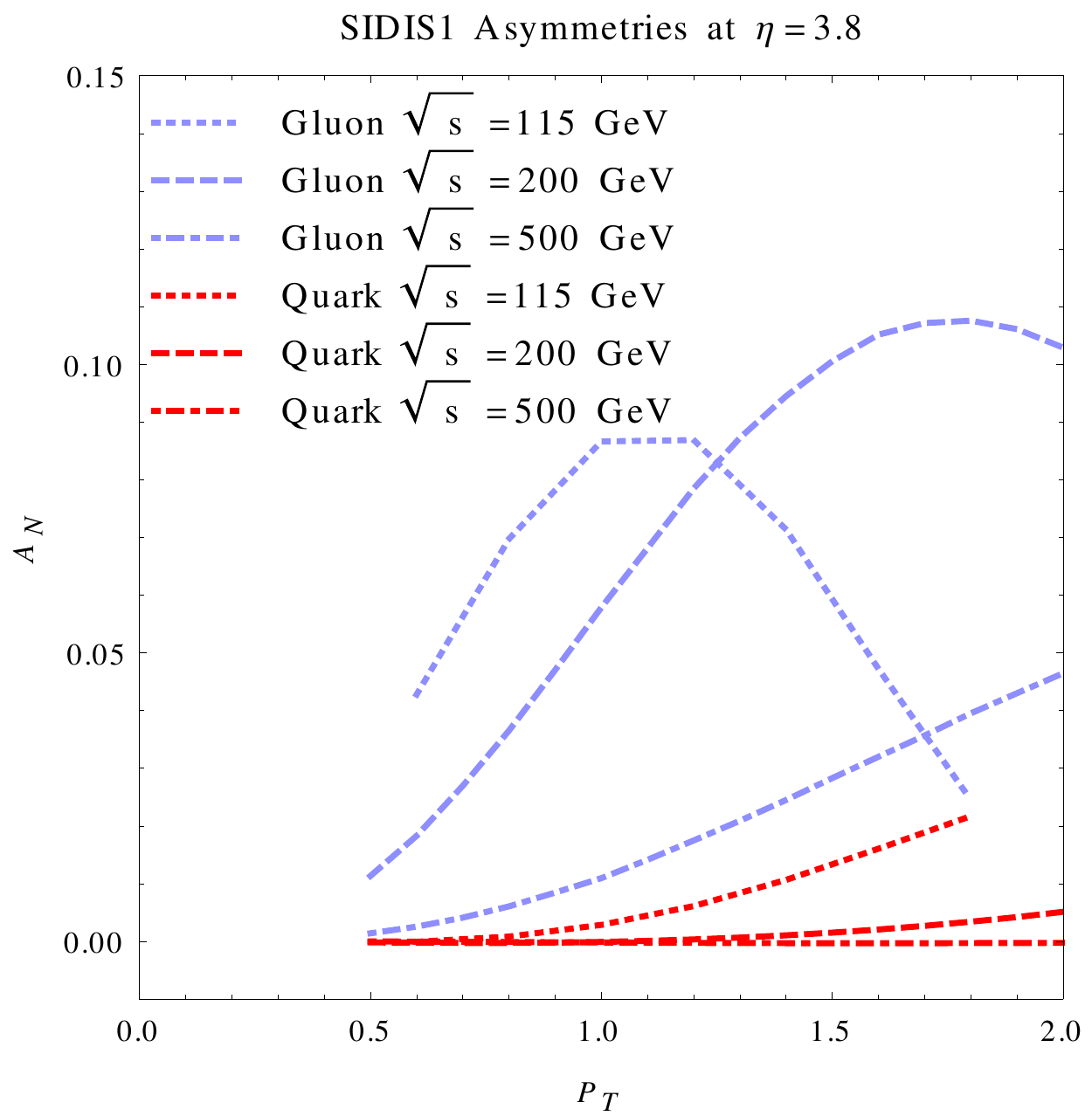}}
\hspace*{0.1cm}
\subfigure[]{\includegraphics[width=0.5\linewidth]{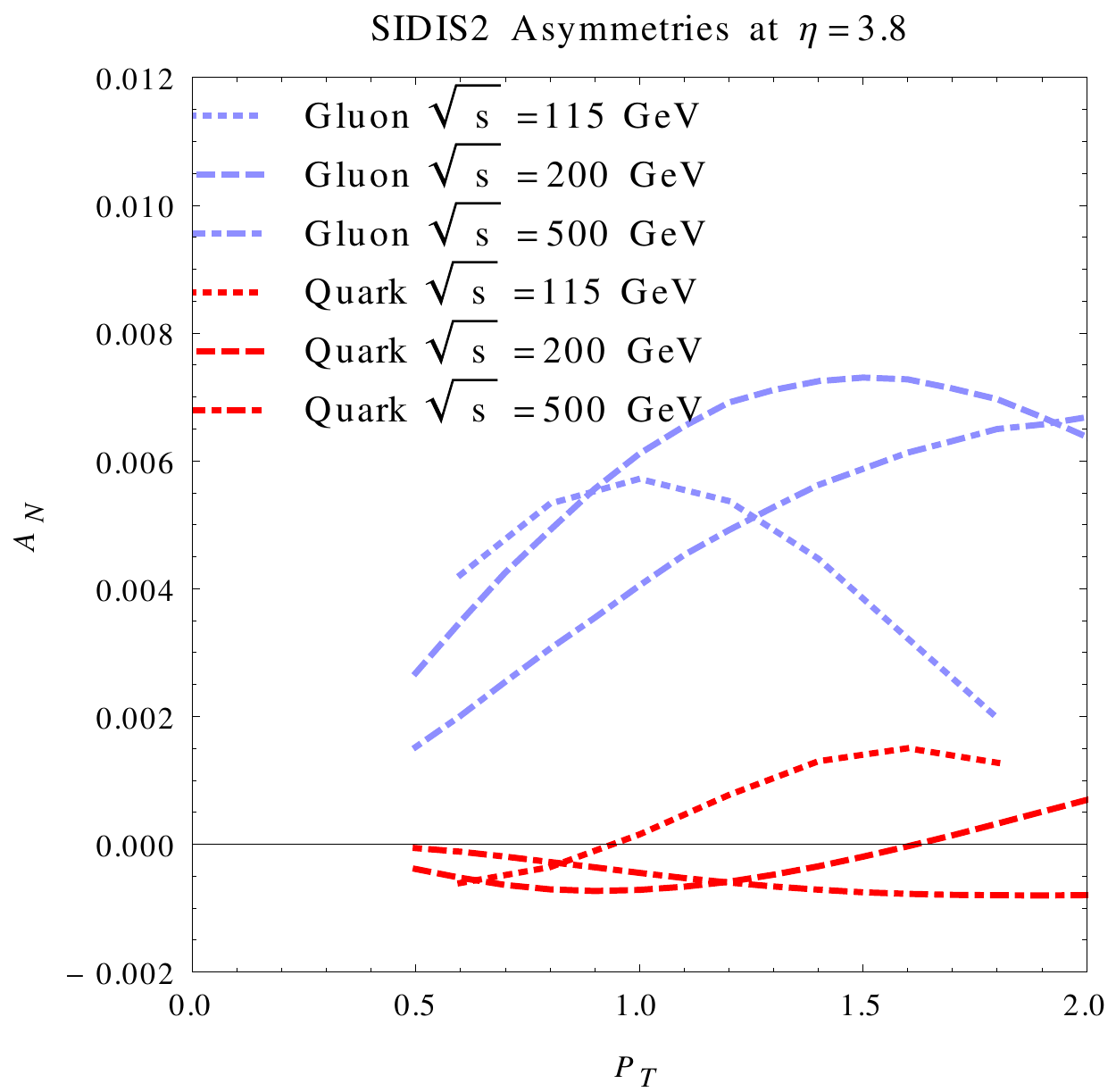}}
\caption{Predictions for gluon and quark contributions to the asymmetry at different values of $\sqrt{s}$ (115 GeV dotted; 200 GeV  dashed; 500 GeV  dot-dashed) and fixed $\eta=3.8$, obtained using (a) (left) SIDIS1 fits and (b) (right) SIDIS2 fits. Gluon contribution is in blue and quark contribution is in red.} 
% Figure 5:  COM dependence of SIDIS1&2.
\label{as_com_dependence}
\end{figure}

Keeping in mind planned experiments \cite{Rakotozafindrabe:2013au, Lansberg:2016urh,Aschenauer:2015eha}, we briefly compare our results at $\sqrt{s}=200$ GeV, with those at $\sqrt{s} = 115$ and 500 GeV. In Fig. \ref{as_com_dependence} we show the results at $\eta=3.8$ (where gluon dominance of the asymmetry is highest), obtained with the SIDIS1 (Fig. \ref{as_com_dependence}(a)) and SIDIS2 (Fig. \ref{as_com_dependence}(b)) Sivers functions, with the gluon and quark contributions shown separately. As can be expected, the asymmetry peaks shift towards higher values of $P_T$ with increasing $\sqrt{s}$.  The relative variation with $\sqrt{s}$ of the asymmetry values for a given $\eta$ and $P_T$ is larger for SIDIS1 than SIDIS2 due to their different $x$-dependence. At lower values of $P_T$($\lesssim 1.5$ GeV), gluon dominance of the asymmetry is in general better at the two higher $\sqrt{s}$ values for both SIDIS1 and SIDIS2, with the gluon dominance increasing faster for SIDIS1. For both the fits, in the region $0.5\text{ GeV}\leq P_T\leq1.5\text{ GeV}$, the asymmetries at $\sqrt{s}=200$ GeV seem to have the best trade off of magnitude and gluon dominance. Of course, the production cross-section is also a crucial factor in determining how well the asymmetry can be measured and it varies by upto 3 orders of magnitude for the given $\sqrt{s}$ and $P_T$ range considered, being highest at 500 GeV.  Therefore measurements of the asymmetry at various centre of mass energies will help get the best understanding of the gluon Sivers function.

In order to better demonstrate the $\sqrt{s}$ dependence and the relative sizes of the gluon and quark contributions to the asymmetry, in Table \ref{com_dependence}, we present the asymmetry predictions integrated over the region $0.5\text{ GeV}\leq P_T\leq2.0\text{ GeV}$, $1.0\leq\eta\leq3.8$. The table shows the integrated contribution of the gluon and quark Sivers functions separately for both the SIDIS1 and SIDIS2 fits. On the last column we give the approximate statistical uncertainty with which the asymmetry can be measured.  As mentioned before, for small asymmetry values this quantity can be written as $\Delta A_N\approx 1/\sqrt{N}$.

The values for the statistical error were obtained assuming a small asymmetry, a branching ratio of $D$ to muons, $BR(D\rightarrow\mu+X)$, which is known to be about 6.7\% \cite{Agashe:2014kda}, a beam polarization of 60\% and an integrated luminosity of 1 $\text{fb}^{-1}$.
\begin{table}[h]
\centering
\begin{tabular}{|l|l|l|l|l|l|l|}
\hline
$\sqrt{s}\small\text{ GeV}$ & $\sigma_\text{total}\small\text{ mb}$ & ${A_N}^{SIDIS1}_\text{gluon}$ & ${A_N}^{SIDIS1}_\text{quark}$ & ${A_N}^{SIDIS2}_\text{gluon}$ & ${A_N}^{SIDIS2}_\text{quark}$ & $\Delta A_N$ (statistical) \\ \hline
115                         & $3.1\times10^{-3}$                    & $5\times10^{-2}$              & -$6.7\times10^{-4}$           & $8.6\times10^{-3}$            & -$1.4\times10^{-3}$           & $3\times10^{-3}$           \\ \hline
200                         & $8.6\times10^{-3}$                    & $3.5\times10^{-2}$            & -$5.5\times10^{-4}$           & $7.3\times10^{-3}$            & -$8.5\times10^{-4}$           & $1.8\times10^{-3}$         \\ \hline
500                         & $3\times10^{-2}$                      & $1.4\times10^{-2}$            & -$2.5\times10^{-4}$           & $5.4\times10^{-3}$            & -$3.3\times10^{-4}$           & $1\times10^{-3}$           \\ \hline
\end{tabular}
\caption{Integrated asymmetries ($0.5\text{ GeV}\leq P_T\leq2.0\text{ GeV}$, $1.0\leq\eta\leq3.8$) with the SIDIS1 and SIDIS2 Sivers functions at different c.o.m energies. Gluon and quark contributions listed separately. These asymmetry values are much smaller than those shown in the differential plots since the integration region includes low values of $\eta$ where the asymmetries are smaller.}
\label{com_dependence}
\end{table}

We find  that the integrated asymmetries decrease with increasing $\sqrt{s}$ in the range considered, but are generally in the same order of magnitude. For the SIDIS1 fit, the gluon contribution to the asymmetry clearly dominates, being almost two orders of magnitude larger than the quark constribution. For the SIDIS2 fit, the quark contribution to the integrated asymmetry is larger, being upto 16\% of the gluon value at $\sqrt{s}=115$ GeV.

\section{Conclusions}

We have presented in this paper, Single Spin Asymmetry predictions for $D$-meson production in hadronic collisions. For this purpose, we use a generalized parton model (GPM) approach and different available fits and models of the gluon Sivers function. We also studied the effect of TMD evolution of the densities on the asymmetry predictions. We presented results at $\sqrt{s}=200$ GeV for the range $-0.7\leq x_F\leq0.7$ at a fixed $P_T=1.5$ GeV and the range $0.5\text{ GeV}\leq P_T\leq3.5\text{ GeV}$ at fixed pseudorapidity values $\eta=2.0\text{ and }3.8$.  We also studied the dependence of the asymmetries on the centre of mass energy by considering the asymmetries from the DMP fits at two other c.o.m values $\sqrt{s}=$115 GeV and 500 GeV.

We find that the SIDIS1 fit of the gluon Sivers function gives sizeable asymmetries in the regions $0.1\lesssim x_F \lesssim 0.7$ with $P_T=1.5\text{ GeV}$ and $1.0\lesssim P_T \lesssim3.0$ with pseudorapidity $\eta=3.8$. The SIDIS2 fit of the gluon Sivers function gives much smaller estimates of the asymmetry in these kinematic ranges. Nevertheless, it is still non-negligible and dominates over the quark contribution. Asymmetries from both fits in the less forward region of $\eta=2.0$ are smaller than at $\eta=3.8$, but the cross-section is larger by an order of magnitude or more, hence the asymmetries may be more easily measurable.  On the other hand the asymmetry values at $\eta=3.8$ are larger by almost an order of magnitude in the low $P_T$ region, hence it may be worthwhile measuring the asymmetry at both low and high pseudorapidity. Overall, peak asymmetry predictions from the two fits are in the range $0.5\%\lesssim|A_N|\lesssim11\%$. These values are large enough to be measurable at RHIC. Note that while we accounted for the brancing ratio $BR(D\rightarrow\mu+X)$ in evaluating the statistical uncertainty in the asymmetry measurement, the kinematics of the decay was not investigated. In fact such a study taking into account the planned acceptance for muons, $1.0\leq\eta\leq4.0$ \cite{MingLiu}, will be very interesting and is under progress.

We find that the inclusion of TMD evolution causes overall asymmetry predictions to diminsh. The peak asymmetry prediction obtained with a maximal gluon Sivers function and TMD evolved densities goes down to less than a third of the peak value obtained with DGLAP evolved densities. The predictions obtained with the BV models, where the gluon Sivers function is modelled upon the quark Sivers functions, also display similar behaviour with the peak asymmetry values dropping by upto a fifth. In general, the effect of TMD evolution on asymmetries found here is qualitatively similar to our earlier results on the electroproduction of  $J/\psi$. 

Currently, all the information we have about the GSF is from fits whose applicability in various processes may be affected by issues of factorisation, and validity of the assumptions involved in rather simple models. This highlights the importance of identifying probes which are highly sensitive to the gluon Sivers function from a variety of processes. A complete understanding of the correlation between the proton spin and gluon transverse momentum can be achieved only upon studying the effective gluon Sivers functions in various processes while taking into account its process dependence. 

The production of heavy flavours - bound states like $J/\psi$ \cite{Godbole:2013bca, Godbole:2014tha, Yuan:2008vn} and heavy mesons like $D$-mesons \cite{Anselmino:2004nk, Kang:2008ih} hold the potential of giving ``clean" probes of the gluon Sivers function. It may be noted that the predictions for charmonium production in hadronic collisions suffer from somewhat large uncertainties due to the lack of clarity on the model that correctly describes all the currently available data on charmonium production \cite{Yuan:2008vn}. On the other hand,  $D$-meson production involves the unknown fragmentation functions, which however, are determined well from fits to data \cite[and references therein]{Cacciari:1996wr, Agashe:2014kda}.  It would also be interesting to set up a framework so that these predictions for heavy flavour production obtained in the generalised parton model can be directly compared with the expectations in the twist-3 formalism \cite{Kang:2008ih} in the region of their overlap.

\section{Acknowledgements}

R.M.G. wishes to acknowledge support from the Department of Science and
Technology, India under Grant No. SR/S2/JCB-64/2007 under the J.C. Bose Fellowship scheme.  A.M would like to thank the Department of Atomic Energy-BRNS, India, for financial support under Grant No. 2010/37P/47/BRNS and Department of Science and Technology, India for financial support under Grant No.EMR/2014/0000486 . AM would also like to thank  CHEP, IISc, Bangalore for their kind hospitality. We would like to thank Dr. Asmita Mukherjee for bringing Ref. \cite{Anselmino:2004nk} to our attention. 

\section{Appendix}

\subsection{Meson production kinematics}
D'Alesio and Murgia have worked out the kinematics relating the observed hadron momentum in the lab to the momenta of the partons involved in the process \cite{D'Alesio:2004up}. We outline it below for clarity. The only addition is the solution for the on-shell condition in terms of the fragmentation variable $z$ for the heavy quark case.

In the following, all momenta are given in the proton-proton c.o.m frame with the polarized proton `A' moving along the positive $Z$ and the unpolarized proton `B' moving along the negative $Z$ axis. By convention, the polarization of proton `A' is chosen to be along the $Y$ axis and the $D$ meson production plane is taken to be XZ.

 Hence the momenta of the protons and the $D$-meson momentum can be written as
\be
 P_A=\frac{\sqrt{s}}{2}(1,0,0,1), P_B=\frac{\sqrt{s}}{2}(1,0,0,-1)\text{ and } P_D=(E_D,P_T,0,P_L)
\ee
where the masses of the protons have been neglected.

The massless partons `a' and `b' inside protons `A' and `B' are described by light-cone momentum fractions $x_a=P_a^+/P_A^+$, $x_b=P_b^-/P_B^-$ and transverse momenta $\mathbf{k}_a$ and $\mathbf{k}_b$ respectively. Their momenta are given by,
\bea
P_a&=x_a\frac{\sqrt{s}}{2}\left(1+\frac{k_{\perp a}^2}{x_a^2 s},\frac{2k_{\perp a}}{x_a\sqrt{s}}\cos\phi_a,\frac{2k_{\perp a}}{x_a\sqrt{s}}\sin\phi_a,1-\frac{k_{\perp a}^2}{x_a^2 s}\right)
\\\nonumber
P_b&=x_b\frac{\sqrt{s}}{2}\left(1+\frac{k_{\perp b}^2}{x_b^2 s},\frac{2k_{\perp b}}{x_b\sqrt{s}}\cos\phi_b,\frac{2k_{\perp b}}{x_a\sqrt{s}}\sin\phi_b,-1+\frac{k_{\perp b}^2}{x_b^2 s}\right)
\eea
where $\phi_a$ and $\phi_b$ are the azimuthal angles of partons `a' and `b' respectively.

These two partons produce a heavy parton `c' (which further fragments into the heavy meson) through the process $ab\rightarrow c\bar{c}$. The momentum of the parton `c' is described by `$z$', the light-cone momentum fraction of the heavy meson and $\mathbf{k}_D$, the transverse momentum of the meson with respect to the parton `c'.

The $D$-meson three-momentum $\bf p_D$ can be split into a component along the three-momentum of the fragmenting heavy quark, $\bf p_c$, and one perpendicular to it. 
Rotating to a frame where $\bf p_c$ is along the z-axis, the meson momentum is,
\be
P_D=(E_D,0,0,|\mathbf{p}_D - \mathbf{k}_D|) +  (0,\mathbf{k}_D)
\ee
where $P_D$ has been split into longitudinal and perpendicular components as mentioned above. In this frame, $\mathbf{k}_D$ is simply $(k_{D_x},k_{D_y},0)=(\mathbf{k}_{D_\perp},0)$. In the lab frame however, $\mathbf{k}_D$ can have all three components non-zero and is specified as,
\be
\mathbf{k}_D=k_D(\sin\theta\cos\phi,\sin\theta\sin\phi,\cos\theta) \text{, with } |\mathbf{k}_D|=|\mathbf{k}_{D_\perp}|
\ee
and the orthogonality condition $\mathbf{k}_D.\mathbf{p}_c=0$ ensures that $\mathbf{k}_D$ lies in a plane perpendicular to $\mathbf{p}_c$. The light-cone momentum fraction $z$ is given by,
\be
z=\frac{P_D^+}{P_c^+}=\frac{E_D+|\mathbf{p}_D-\mathbf{k}_D|}{E_c+|\mathbf{p}_c|}=\frac{E_D+\sqrt{\mathbf{p}_D^2-\mathbf{k}_D^2}}{E_c+\sqrt{E_c^2-m_c^2}}
\label{light-cone-z}
\ee
From the above equation, one obtains for the energy of the fragmenting parton,
\be
E_c=\frac{m_c^2+\left((E_D+\sqrt{\mathbf{p}_D^2-\mathbf{k}_D^2})/z\right)^2}{2\left((E_D+\sqrt{\mathbf{p}_D^2-\mathbf{k}_D^2})/z\right)}
\label{energy-relation}
\ee

The expression for $\mathbf{p}_c$ can be obtained from the fact that it is collinear with $\mathbf{p}_D - \mathbf{k}_D$ and that the unit vector constructed out of both must therefore be equal,
\be
\vec P_c=\sqrt{E_c^2-m_c^2}\frac{\mathbf{p}_D-\mathbf{k}_D}{|\mathbf{p}_D-\mathbf{k}_D|}
\label{momentum-relation}
\ee

Eqs. (\ref{energy-relation}) and (\ref{momentum-relation}) relate the energy and momentum of the observed $D$-meson with that of the fragmenting parton for given values of $k_D$ and $z$. 

The term $d^3 \mathbf{k}_D \, 
\delta (\mathbf{k}_D \cdot \hat{\mathbf{p}}_c)$ in Eqs. (\ref{final-ssa}) and (\ref{final-unp}) ensures that the $\mathbf{k}_D$ integration is only over momenta transverse to the fragmenting parton:
\be
d^2\mathbf{k}_{D_\perp}=d^3 \mathbf{k}_D \, 
\delta (\mathbf{k}_D \cdot \hat{\mathbf{p}}_c)=dk_D\text{ }k_D\text{ }d\theta \text{ }d\phi\frac{|\mathbf{p}_D-\mathbf{k}_D|}{P_T\sin\theta\sin\phi_1}\left[\delta(\phi-\phi_1)+\delta(\phi-(2\pi-\phi_1))\right]
\ee
where,
\be
\cos \phi_1=\frac{k_D-P_L\cos\theta}{P_T\sin\theta}
\ee
Limits on $|\mathbf{k}_D|$ can be obtained by requiring $|\cos \phi_1|\leq1$. They are,
\be
\text{min}\left[P_L\cos\theta-P_T\sin\theta,0\right]\leq |\mathbf{k}_D| \leq\text{min}\left[P_L\cos\theta+P_T\sin\theta,0\right]
\ee

\subsection{Solving the on-shell condition for $z$}
We give the expressions for the $a_i$ in Eq. \ref{onshellai} below:
\bea
a_1&=&-\left(\frac{1}{2\sqrt{s}}\left(\frac{k_{\perp a}^2}{x_a}+\frac{k_{\perp b}^2}{x_b}\right)+\frac{\sqrt{s}}{2}(x_a+x_b)\right)(E_D+|\mathbf{p}_D-\mathbf{k}_D|)
\\ \nonumber
a_2&=&\frac{m_c^2}{(E_D+|\mathbf{p}_D-\mathbf{k}_D|)^2}\times a_1
\\ \nonumber
a_3&=&\frac{P_L-k_D\cos\theta}{|\mathbf{p}_D-\mathbf{k}_D|}   \left(-\frac{1}{\sqrt{s}}\left(\frac{k_{\perp a}^2}{x_a}-\frac{k_{\perp b}^2}{x_b}\right) + \sqrt{s}(x_a+x_b)\right) \\ \nonumber
& +& \frac{2(P_T-k_D\sin\theta\cos\phi)}{|\mathbf{p}_D-\mathbf{k}_D|}(k_{\perp a}\cos\phi_a+k_{\perp b}\cos\phi_b)-\frac{2k_D\sin\theta\sin\phi}{|\mathbf{p}_D-\mathbf{k}_D|}(k_{\perp a}\sin\phi_a+k_{\perp b}\sin
\phi_b)
\\ \nonumber
a_4&=&m_c^2
\\ \nonumber
a_5&=&(E_D+|\mathbf{p}_D-\mathbf{k}_D|)^2
\\ \nonumber
a_6&=&\hat s
\eea

In terms of these factors, the solution for $z$ is given by,
\bea
z&=&-\frac{a_3 a_4 \sqrt{a_5} a_6}{\sqrt{\frac{a_4^4(-4a_1^2+a_3^2 a_5)^2}{a_5^2}}}
-
\frac{2 a_1 a_5 a_6}{4a_1^2 a_4 - a_3^2 a_4 a_5}
\\ \nonumber
&-&
\sqrt
{
\frac
{a_5 \left(-16a_1^4 a_4^2+a_3^4 a_4^2 a_5^2+4a_1 a_3 a_5^{3/2} \sqrt{\frac{a_4^4(-4a_1^2+a_3^2 a_5)^2}{a_5^2}}\right)(-4a_1^2 a_4+a_5 (a_3^2a_4+a_6^2))}
{a_4^4(-4a_1^2+a_3^2 a_5)^3}
}
\eea
where we have used the simplification $a_2=\frac{a_1 a_4}{a_5}$.

\subsection{Quark Sivers function parameters used for the Boer-Vogelsang (BV) Models}
Here we give the values for the quark Sivers function parameters used in the BV models of the GSF \cite{Boer-PRD69(2004)094025}. The quark Sivers function parameters used in the SIDIS1 and SIDIS2 fits are from Refs. \cite{Anselmino:2005ea} and \cite{Anselmino:2008sga} respectively. The quark Sivers function parameters for the TMD evolved case are from Ref. \cite{Anselmino:2012aa}.

\begin{table}[H]
\centering
\begin{tabular}{l|l|l|l|l|l|l|l|}
\cline{2-8}
 & $N_u$ & $\alpha_u$ & $\beta_u$ & $N_d$ & $\alpha_d$ & $\beta_d$ & $M_1^2$ \\ \hline
\multicolumn{1}{|l|}{SIDIS1} & 0.32 & 0.29 & 0.53 & -1.0 & 1.16 & 3.77 & 0.32 \\ \hline
\multicolumn{1}{|l|}{SIDIS2} & 0.35 & 0.73 & 3.46 & -0.9 & 1.08 & 3.46 & 0.34 \\ \hline
\multicolumn{1}{|l|}{TMD evolved} & 0.75 & 0.82 & 4.0 & -1.0 & 1.36 & 4.0 & 0.34 \\ \hline
\end{tabular}
\caption{Quark Sivers function parameters}
\label{my-label}
\end{table}

\end{document}